\documentclass[12pt,titlepage]{utarticle}
\usepackage{amsmath,amssymb,graphicx,color}
\usepackage[curve]{xypic}
\usepackage{hyperref}

\numberwithin{equation}{section}

\newcommand{\pxpy}[2]{\frac{\partial{#1}}{\partial{#2}}}

\def\BZ{\mathbb{Z}}
\def\BC{\mathbb{C}}

\newcommand{\MR}[1]{{\mathbb{R}^{#1}}}            
\newcommand{\MC}[1]{{\mathbb{C}^{#1}}}            
\newcommand{\MS}[1]{{{\bf S}^{#1}}}               
\newcommand{\PP}[1]{{\mathbf{P}^{#1}}}

\def\Hom{\mathrm{Hom}}
\def\dHom{\mbox{-Hom}}
\def\Ext{\mathrm{Ext}}
\def\dExt{\mbox{-Ext}}

\def\dim{\mathrm{dim\ }}
\def\mcO{\mathcal{O}}
\def\mcA{\mathcal{A}}
\def\mcD{\mathcal{D}}
\def\mcF{\mathcal{F}}
\def\mcG{\mathcal{G}}
\def\mcT{\mathcal{T}}

\def\Ten{\mathrm{Ten}}

\def\ie{\textit{i.e.,\ }}

\newtheorem{conj}{Conjecture}

\begin{document}
\preprint{UTTG--02--05\\
\texttt{hep-th/0502105}\\}

\title{Undoing Orbifold Quivers}

\author{Aaron Bergman}

\oneaddress{Theory Group, Physics Department\\
             University of Texas at Austin\\
             Austin, TX 78712\\ {~}\\
             \email{abergman@physics.utexas.edu}}

\Abstract{A number of new papers have greatly elucidated the derivation of
quiver gauge theories from D-branes at a singularity. A complete story has
now been developed for the total space of the canonical line bundle over
a smooth Fano 2-fold. In the context of the AdS/CFT conjecture, this corresponds
to eight of the ten regular Sasaki-Einstein 5-folds. Interestingly,
the two remaining spaces are among the earliest examples, the sphere and
$T^{11}$. I show how to obtain the (well-known) quivers for these
theories by interpreting the canonical line bundle as the resolution of
an orbifold using the McKay correspondence. I then obtain the correct
quivers by undoing the orbifold. I also conjecture, in general, an autoequivalence
that implements the orbifold group action on the derived cateory. This yields
a new order two autoequivalence for the $\BZ_2$ quotient of the conifold.}

\maketitle
\newpage

\section{Introduction}\label{sec:intro}

The nature of singularities has long been one of the central interesting
questions in string theory. One of the best techniques we have for
understanding this has been to probe them with D-branes. Thus, it is
of great interest to understand the gauge theory that lives on a D-brane
situated at a singularity. This is particularly important in the context
of the AdS/CFT conjecture wherein the most interesting gauge theories
are obtained in such a manner.

While there have been a number of techniques used to derive such
gauge theories, the most powerful at present seems to be that of
exceptional collections in derived categories. First referenced in
\cite{Cachazo:2001sg} and initially developed by \cite{Wijnholt:2002qz},
this technique has
been used to derive quivers for the gauge theories resulting from
all collapsing del Pezzo surfaces in Calabi-Yaus. The procedure in
those references has now been substantially elucidated in the works of
\cite{Herzog:2004qw,Aspinwall:2004vm,BridgeTStruct}. In particular, the last reference
makes concrete most of the relevant mathematics. While the issues of
superpotentials is still not completely worked out, we will see that in
many cases we can use physical insight to guess the correct answer.

In the AdS/CFT conjecture, we take a stack of D3-branes on $\MR{4}$
located at the tip of a 6 (real-)dimensional cone 
with a Calabi-Yau metric of the form
\begin{equation}
\label{CYmetric}
ds^2 = dr^2 + r^2 ds^2_{M^5}\ .
\end{equation}
When we take the near horizon limit of such a metric, it is easy to see
that the resulting geometry is $AdS_5 \times M^5$. An interesting feature
of the metric (\ref{CYmetric}) is that the radial coordinate $r$ does not
have to extend to zero. Any CY metric of this form defines a Sasaki-Einstein
(SE) structure on $M^5$. A characteristic feature of such metrics is
that they have a nondegenerate vector field that can be integrated to
a foliation of the 5-fold. These foliations can be divided into three
different types: (1) \textit{regular}, where all the leaves have the same
lengths; (2) \textit{quasiregular}, where the leaves can have different
lengths and (3) \textit{irregular}, where the leaves are noncompact. In
case (1), the space of leaves is a Fano\footnote{Fano means that the
the anticanonical line bundle is ample (positive).} K\"ahler-Einstein (KE) 2-fold,
and in case (2) it is a Fano K\"ahler-Einstein orbifold. The last case
remains mysterious, although some examples have been recently investigated
\cite{Gauntlett:2004yd,Gauntlett:2004hh,Benvenuti:2004wx,Herzog:2004tr,Hashimoto:2004ks,
Benvenuti:2004dy,Bertolini:2004xf,Martelli:2004wu,Benvenuti:2005wi}. For the remainder of this paper, we will deal only with the first case.

In both the first two cases, the 5-fold is the total space of a circle
(V-)bundle over the base.  All smooth Fano, KE 2-folds are known. They are the
third through eighth del Pezzo surfaces\footnote{The notation
$dP_n$ refers to $\PP{2}$ blown up at $n$ points.}, $dP_3,\dots,dP_8, \PP{2}$
and $\PP{1} \times \PP{1}$. All regular SE 5-folds are circle bundles
over these spaces. It was shown by Friedrich and Kath \cite{FriedKath} that the
Euler class of these must integrally divide the canonical class
of the base. This
gives the following complete classification. First, we introduce the
\textit{Fano index} of a Fano variety, $X$. This is the largest natural number,
$I$, such that $c_1(X) / I \in H^2(X,\BZ)$. The Fano index of all the
del Pezzos is 1, so the only SE space is the total space of the circle
bundle corresponding to the canonical class. For $\PP{2}$, the Fano
index is 3, so we have circle bundles corresponding to the canonical
class and the canonical class divided by three. These manifolds are
easily seen to be $\MS{5}/\BZ_3$ and $\MS{5}$ respectively. Finally,
the Fano index of $\PP{1}\times\PP{1}$ is 2 and the corresponding 
5-folds are denoted $T^{11}/\BZ_2$ and $T^{11}$. In all these situations,
the cone given by the metric (\ref{CYmetric}) can be thought of as
`filling in' the circle bundle, \ie as the total space of the line bundle
with same first Chern class as the Euler class of the circle
bundle. As noted earlier, the metric may not extend to the zero
section of this line bundle.

While all these SE 5-folds are relevant for the AdS/CFT conjecture,
the techniques given in \cite{Herzog:2004qw,Aspinwall:2004vm,BridgeTStruct}
only work for the case of the canonical bundle. The reason for this is that,
in order to work at large volume, we must be able to blow up the base
of the cone. However, the CY metric only extends to the zero
section when the
canonical class of the cone is trivial which implies that the bundle must
be the total space of the canonical bundle of the base. However, all
hope is not lost for the other two cases. In each case, the singularity we
obtain when we collapse the `tip' of the cone to a point has an alternate
resolution. For $T^{11}$, the cone is the conifold with its well-known
resolution to $\mcO(-1)\oplus\mcO(-1) \rightarrow \PP{1}$. For $S^5$,
the cone is just $\BC^3$ which needs no resolution.

While it is possible to analyze these two cases in terms of their
CY resolutions, we will go a different route in this paper in the hope
that the techniques developed will eventually be applicable to the less
well-understood quasiregular case. We will exploit the fact that the
canonical \textit{circle} bundle is always an orbifold when the Fano
index is greater than one. This suggests that we can interpret the
quivers obtained by the techniques of
\cite{Herzog:2004qw,Aspinwall:2004vm,BridgeTStruct} as orbifolds of the
quivers we desire. We will show how this can be accomplished and show
that it obtains the correct answers for $\MS{5}$ and $T^{11}$.

The paper is organized as follows. In section \ref{sec:helices},
we introduce exceptional collections and helices and show how to
obtain quiver gauge theories from them. We make a conjecture as
to the form of the superpotential. In section
\ref{sec:algo}, we present an algorithm to undo the orbifold of
the quiver, and we show that it gives the correct results.
In section \ref{sec:cstar}, we use the
simpler example of $\BC^*$ bundles to motivate the algorithm and
to introduce some orbifold techniques. In section \ref{sec:mckay},
we show how we can use the McKay correspondence to treat the derived
category of coherent sheaves on the total space of the canonical bundle
as an equivariant derived category. In section \ref{sec:monod},
we conjecture a Fourier-Mukai transform that implements the action of
the orbifold group on the derived category and show how this
justifies the algorithm presented in section \ref{sec:algo}.

\section{Exceptional collections, helices and quivers}\label{sec:helices}

We will assume familiarity with the language of derived categories.
Excellent references on the subject are Aspinwall's review
\cite{Aspinwall:2004jr} and the textbooks \cite{Weibel,GelMan}. Because
exceptional collections are well established in this area, we will
be brief in our review. For more details, see \cite{Herzog:2004qw,Aspinwall:2004vm}
and references therein.

\subsection{Exceptional collections and quivers}

An exceptional object, $E$, in a derived category\footnote{Really, any
triangulated category.} satisfies the following identities:
\begin{equation}
\begin{gathered}
 \Hom(E,E) = \mathbb{C}\ , \\
 \Hom(E,E[k]) = \Ext^k(E,E) = 0\quad \mathrm{for}\quad k \neq 0\ .  
\end{gathered}
\end{equation}

An exceptional collection is an ordered collection of exceptional
objects, $E_i$, $i = 0\dots n-1$, such that
\begin{equation}
i>j \Rightarrow \Ext^k(E_i,E_j) = 0\quad \forall k \in \BZ{}\ .
\end{equation}

An exceptional collection is called \textit{full} if it generates the
triangulated category and is called \textit{strong} if
\begin{equation}
\Ext^k(E_i,E_j) = 0 \quad \forall k \neq 0 \quad \forall i,j \ .
\end{equation}

It was shown by Bondal \cite{BondalQuiv} that, given the data of an
full, strong exceptional collection of objects in
a triangulated category, we can construct
a quiver such that the derived category of representations of
the quiver is equivalent to the original triangulated category.
For our purposes, the original triangulated category will always
be the derived category of coherent sheaves on some variety, $X$.
Thus, we have an equivalence of triangulated categories
$\mcD(X) \cong \mcD(A - \mathrm{Mod})$ where we identify representations
of the quiver with modules of the quiver algebra, $A$.

The construction proceeds as follows. Given a full, strong
exceptional collection, let us form the object
\begin{equation}
\label{smallT}
T = \bigoplus_{i=0}^{n-1} E_i\ .
\end{equation}
Then, the endomorphism algebra of this object is the algebra of a
quiver with relations. In fact, it will be more useful to look at
the opposite algebra\footnote{This essentially interchanges the
roles of left and right modules.}, so we define
\begin{equation}
A^\mathrm{op} = \Hom(T,T)\ .
\end{equation}

We will assume familiarity with the theory of representations of
quivers.\footnote{For an extensive reference, see \cite{Derksen}. For the
relevant parts, see \cite{Aspinwall:2004vm}.} Recall that there are two
types of distinguished representations of a quiver associated to
any given node. First, there is the simple representation, $S_i$,
given by a one dimensional vector space at that node with all maps
equal to zero. Second,
there is the projective representation, $P_i$,  where, associated to the
node labelled $j$, we have a vector space with basis given by the
set of paths from node $i$ to node $j$. The maps are given by concatenation.
With these definitions, we have the identification
\begin{equation}
\Hom(E_i,E_j) \cong \Hom(P_i,P_j) = \mbox{paths from $j$ to $i$}\ .
\end{equation}


\begin{figure}
\centerline{
\xymatrix{
 & {\mathcal{O}(1)} \ar@3{<-}[]-/u 4mm/*{\bullet};
[dr]-/r 6mm/*{\bullet}^{b_j} \\
*{\mathcal{O}} \ar@3{<-}[]-/l 3mm/*{\bullet};
[ur]-/u 4mm/^{a_i} & & *{\mathcal{O}(2)}
}}
\caption{The quiver for $\PP{2}$.}
\label{p2quiver}
\end{figure}

Let us now work out the quivers in the two examples that will be
the focus of this paper. First, for $\PP{2}$, a full, strong exceptional
collection is $(\mcO,\mcO(1),\mcO(2))$. It is easy to compute
\begin{equation}
\label{p2homs}
\begin{gathered}
\Hom(\mcO,\mcO(1)) = \Hom(\mcO(1),\mcO(2)) = \BC^3\ , \\
\Hom(\mcO,\mcO(2)) = \BC^6\ .
\end{gathered}
\end{equation}
This tells us that the quiver is the one given in figure
\ref{p2quiver}. To determine the relations, we note that the maps
in equation (\ref{p2homs}) are simply multiplication by sections
of $\mcO(1)$. The relations come from the fact that this
multiplication is commutative, \ie $xy = yx \in H^0(\mcO(2))$ for
$x,y\in H^0(\mcO(1))$. Thus, using the labels in the diagram, we
have the relations $a_i b_j = a_j b_i$, $i,j=1,2,3$. This can be
succinctly written as $\epsilon^{ijk} a_j b_k = 0$.

\begin{figure}
\centerline{
\xymatrix{
{} & {\mathcal{O}(0,1)} \ar@2{<-}[]-/u 4mm/*{\bullet};[dr]-/r
8mm/*{\bullet}^{b_j} & {}\\
{\mathcal{O}} \ar@2{<-}[]-/l 4mm/*{\bullet};[ur]-/u 4mm/^{a_i} &
{} &
{\mathcal{O}(1,1)} \ar@2{<-}[]-/r 8mm/*{\bullet};
[dl]-/d 4mm/*{\bullet}^{c_k} \\
 {} & {\mathcal{O}(1,2)} & {}
}}
\caption{The quiver for $\PP{1} \times \PP{1}$.}
\label{p1p1quiver}
\end{figure}

In the case of $\PP{1} \times \PP{1}$, we will work with the full, strong
exceptional collection
\begin{equation}
\label{p1p1exc}
(\mcO,\mcO(0,1),\mcO(1,1),\mcO(1,2))\ .
\end{equation}
Here, $\mcO(a,b)$ denotes the sheaf $\mcO(a) \boxtimes \mcO(b)$.
It is a short exercise to verify that the quiver is given as in
figure \ref{p1p1quiver} with the relations $a_i b_j c_k = a_k b_j
c_i$, $i,j,k=1,2$. We can summarize this as $\epsilon^{ik} a_i
b_j c_k = 0$.

\subsection{Completing the quiver}

Now that we have quivers corresponding to the 2-folds, we need to
understand how to obtain quivers corresponding to the full cones.
In particular, we will work with the total space of the canonical
line bundle, denoted by $K_X$ for the 2-fold $X$. Following
Aspinwall \cite{Aspinwall:2004vm}, we will describe a somewhat \textit{ad hoc}
procedure for obtaining the correct quiver. Please see the
original paper for more details.

We will show the next section how to obtain a
collection of objects in $\mcD(X)$, $S_i$, dual to the exceptional
collection $E_i$. These
objects correspond to the simple representations of the quiver.
Some useful properties of these representations (see, for
example, \cite{Aspinwall:2004vm}) are
\begin{equation}
\label{extmean}
\begin{gathered}
\dim\Ext^1(S_i,S_j) = n_{ij} \\
\dim\Ext^2(S_i,S_j) = r_{ij} \\
\dim\Ext^3(S_i,S_j) = rr_{ij} \\
\vdots
\end{gathered}\ .
\end{equation}
Here, $n_{ij}$ is the number of arrows (\textit{not} paths)
from node $i$ to node $j$
in the quiver, $r_{ij}$ is the number of relations between paths
that go from node $i$ to node $j$ and $rr_{ij}$ is the number of
relations between relations, \textit{ad infinitum}.

Tensoring with one of these representations adds one to the 
dimension of the vector space at
a given node. In the gauge theory, this corresponds to adding one to the
rank of the gauge group at the node. This allows us to identify these
representations with the  fractional branes in
the string theory. We only must embed them in the cone $K_X$ to
obtain the correct answer. Let $s$ denote the zero section of the
bundle $K_X$. Then, the relevant objects are $s_*(S_i)$.

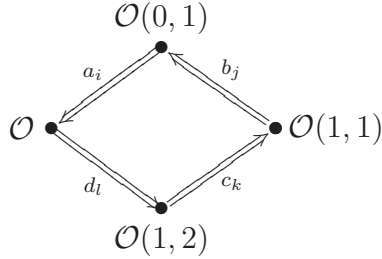
\begin{figure}
\centerline{
\xymatrix{
{} & {\mathcal{O}(0,1)} \ar@2{<-}[]-/u 4mm/*{\bullet};[dr]-/r
8mm/*{\bullet}^{b_j} & {}\\
{\mathcal{O}} \ar@2{<-}[]-/l 4mm/*{\bullet};[ur]-/u 4mm/^{a_i} &
{} &
{\mathcal{O}(1,1)} \ar@2{<-}[]-/r 8mm/*{\bullet};
[dl]-/d 4mm/*{\bullet}^{c_k} \\
 {} & {\mathcal{O}(1,2)} \ar@2{<-}[]-/d 4mm/;[ul]-/l 4mm/^{d_l} &
{}
}}
\caption{The completed quiver for $\PP{1} \times \PP{1}$.}
\label{p1p1cquiv}
\end{figure}

There exists a spectral sequence that gives us the Ext groups
between these objects \cite{Aspinwall:2004vm,Katz:2002gh,SeiThom}. It gives
\begin{equation}
\label{extreln}
\Ext^i_{K_X}(s_*(S_i),s_*(S_j)) = \Ext^i_X(S_i,S_j) \oplus
\Ext^{3-i}_X(S_j,S_i)\ .
\end{equation}
In order to avoid tachyons, we have to work with quivers that do
not have any $\Ext^3$s. What this equation then tells us is that,
for any relation in the quiver for $X$, we must draw a line in the
opposite direction in the quiver for $K_X$. We will call this new
quiver  the `completed quiver'.
Also, for any arrow in the original quiver, we have a relation in the
completed quiver going in the opposite direction. Finally, we have
relations between the relations at every node. The completed
quivers for $\PP{2}$ and $\PP{1}\times\PP{1}$ are given in
figures \ref{p2cquiv} and \ref{p1p1cquiv}.

\begin{figure}
\centerline{
\xymatrix{
 & {\mathcal{O}(1)} \ar@3{<-}[]-/u 4mm/*{\bullet};
[dr]-/r 6mm/*{\bullet}^{b_j} \\
*{\mathcal{O}} \ar@3{<-}[]-/l 3mm/*{\bullet};
[ur]-/u 4mm/^{a_i} & & *{\mathcal{O}(2)}
\ar@3{<-}[]-/r 6mm/;[ll]-/l 3mm/^{c_k}
}}
\caption{The completed quiver for $\PP{2}$.}
\label{p2cquiv}
\end{figure}

It now remains to determine the relations of the completed
quiver. To do this rigorously is difficult, possibly involving higher
products in the algebra of Ext groups.\footnote{Some computations
of superpotentials in this context appear in \cite{Aspinwall:2004bs}.}
We can easily guess a set
of relations that satisfy the aforementioned properties, however.
The key is to remember that relations in quiver gauge theories
correspond to F-terms in the gauge theory and as such are given
as the derivatives of
a superpotential. Let us write the set of relations
of  the quiver for $X$ as $R^a = 0$ where $a$ ranges from $1$ to
$\sum r_{ij}$. Equation (\ref{extreln}) tells us that, for each of
these relations, we add a new arrow to the completed quiver which we will
denote $r_a$. Note that any relation is a sum of paths from a
node $i$ to a node $j$ while the corresponding added arrow
points from $j$ to $i$. Thus, the quantity
\begin{equation}
\label{superpotential}
W = \sum_a r_a R^a
\end{equation}
is gauge invariant in the quiver gauge theory.
This is the superpotential, and the relations for the completed
quiver are derived from it as
\begin{equation}
\pxpy{W}{a_i} = 0
\end{equation}
where $a_i$ ranges over all arrows in the completed quiver. It is easy to
see that this set of relations obeys all the properties implied
by the equation (\ref{extreln}).

Applying this to our examples, we obtain for $\PP{2}$:
\begin{equation}
\label{p2super}
W = \epsilon^{ijk} a_i b_j c_k\ ,
\end{equation}
and for $\PP{1} \times \PP{1}$:
\begin{equation}
\label{p1p1super}
W = \epsilon^{ik}\epsilon^{lj}a_l b_i c_j d_k\ .
\end{equation}
These are the well-known, correct superpotentials for these examples.

\subsection{Mutations and helices}\label{sec:helmut}

In order to justify the preceding manipulations and to identify
the algebra of the completed quiver, we must first introduce the
notions of a mutation and a helix.

There are in fact two notions of a mutation that we will need.
We will begin by describing a mutation in a triangulated
category. Given two objects, $E$ and $F$, we define the left
mutation, $L_EF$, by the triangle
\begin{equation}
L_E F \longrightarrow \Hom(E,F) \otimes E \longrightarrow F
\end{equation}
where the second arrow is the evaluation map. It is not hard to
see that, given an exceptional pair, $(E,F)$, the pair, $(L_E F,
E)$, is also exceptional. This defines a braid group action on
exceptional collections \cite{GoroExc}. We can
similarly define a right mutation, but we will not need it here.

In fact, if the objects $E$ and $F$ are both the images of
coherent sheaves in the derived category, the object $L_E F$ will
often have a single nonzero cohomology sheaf. This can be proven
to be the case\footnote{This is a sufficient but not
necessary condition.} when our variety has no rigid torsion sheaves and
$h^0(-K_X) \ge 2$. As this will always be the case here, let us define a new
object $L^s_E F$, the left sheaf mutation, to be this 
cohomology sheaf. A more proper definition is
given in \cite{GoroMov}.

We can now define the dual objects mentioned in the previous
section. Given an exceptional collection $(E_0,\dots,E_{n})$,
we can define the new collection
\begin{equation}
\label{dualcol}
(F_0,\dots,F_n) = (L^n E_n, L^{n-1} E_{n-1},\dots,L_{E_0}
E_1,E_0)\ .
\end{equation}
The notation $L^n$ refers to $n$ applications of left mutation
(\textit{not} sheaf mutations).
For example, $L^2 E_2 = L_{E_0} L_{E_1} E_2$. These obey
\cite{BridgeTStruct,BondalQuiv}
\begin{equation}
\Ext^k(E_i,F_{n-j}[j]) = \BC \delta_{ij} \delta_{k0}\ .
\end{equation}
This gives us the dual collection\footnote{
In most of the literature, the $F$s are referred to as the dual
collection. However, the $S$s are the important objects for us,
so we prefer to reserve the term dual for them.},
$S_j = F_{n-j}[j]$. For the
collection $(\mcO,\mcO(1),\mcO(2))$ over $\PP{2}$, the dual
collection, $S_i$, is
given by $(\Omega^0,\Omega^1(1)[1],\Omega^2(2)[2])$.
For the collection (\ref{p1p1exc}), the dual collection is
$(\mcO,\mcO(-1,0)[1],\mcO(1,-1)[1],\mcO(0,-1)[2])$.

Also, given an exceptional collection $(E_0,\dots,E_{n})$, we
can take the rightmost element $E_{n}$ and mutate it to
the far left giving the new exceptional collection
$(E_{-1},\dots,E_{n-1})$. It is a theorem of Bondal
\cite{BondalQuiv} that, if the exceptional collection is full,
\ie generates the derived category, then
\begin{equation}
\label{farmutate}
E_{-1} = L^n E_{n} = E_{n} \otimes K[m-n]
\end{equation}
where $m$ is the dimension of the variety that we are working on.

Bondal \cite{BondalQuiv} defines a helix as a collection of objects,
$E_i$, $i=-\infty,\dots,\infty$, such that
\begin{equation}
E_i = E_{i+n} \otimes K[m-n]\ .
\end{equation}
In particular, this defines a helix of length $n+1$. The relation
(\ref{farmutate}) shows that any full exceptional
collection is the basis of a helix that can be formed by left
(and right) mutations.

This, however, does not seem to be the most appropriate
definition of a helix.\footnote{In this context, this fact has also been observed
by Christopher Herzog and Tom Bridgeland to my knowledge.} In fact,
most of the theorems about helices are proven \cite{BondPol} in
the case $m=n$ in the above notation. As we will see, a better
notion of a helix is a collection of \textit{sheaves}, rather
than objects in the derived category, that obey
\begin{equation}
\label{newhelix}
E_i = E_{i+n} \otimes K\ .
\end{equation}
Given an exceptional collection of length $n+1$, we can use the
notion of sheaf mutation to define one of these helices by
$E_{-1} = (L^s)^n E_n$ and the corresponding relation for right
sheaf mutations.  We will use the term helix solely to refer to
a collection of sheaves that obey (\ref{newhelix}). This sort of helix
appears, for example, in \cite{GoroMov}.

Bondal and Polishchuk \cite{BondPol} call a helix geometric if, 
for all $i \le j$, it obeys
\begin{equation}
\Hom(E_i,E_j[k]) = 0 \mathrm{\ unless\ } k = 0\ .
\end{equation}
Bridgeland \cite{BridgeTStruct} calls these helices \textit{simple}, and
we will follow his usage. Note that, while we are applying
Bondal's definition to our modified version of a helix, it still
makes sense. In fact, Bondal and Polishchuk \cite{BondPol} 
show  that, for their definition of a helix, it can only be
geometric/simple
in the case that $m=n$, \ie when it is also a helix under our
definition.

For our two running examples, on $\PP{2}$, we have the simple
helix $E_i = \mcO(i)$ and on $\PP{1} \times \PP{1}$
\begin{equation}
\label{p1p1hel}
\dots,\mcO(-1,-1),\mcO(-1,0),\mcO,\mcO(0,1),\mcO(1,1),\mcO(1,2),
\mcO(2,2),\mcO(2,3),\dots\ .
\end{equation}

\subsection{Deriving the completed quiver}

With these tools in hand, we can proceed to find a quiver algebra
whose derived category of modules is equivalent to the derived
category of the total space $K_X$. This material is from
Bridgeland's paper \cite{BridgeTStruct}.

Given a simple helix $E_i$ generated by an exceptional collection
$(E_0,\dots,E_{n-1})$, we define the following graded algebra:
\begin{equation}
\label{helalg}
\bigoplus_{k\ge 0} \prod_{j-i = k} \Hom(E_i,E_j)\ .
\end{equation}
which Bridgeland calls the \textit{helix algebra}.\footnote{This
is similar to the helix algebra in \cite{BondPol}.} There is a natural
$\BZ$-action given by the isomorphism
\begin{equation}
\otimes K_X : \Hom(E_i,E_j) \longrightarrow(E_{i-n},E_{j-n})\ .
\end{equation}
The invariant subalgebra under this action is called the
\textit{rolled-up helix algebra}. We will denote its opposite
algebra by $B$\footnote{This is the opposite of Bridgeland's
definition, but is consistent with our earlier definition of $A$
and with the conventions of other papers.}.

Now, similar to the object $T$ in (\ref{smallT}), we can define
on $K_X$:
\begin{equation}
\widetilde{T} = \bigoplus_{i=0}^{n-1} \pi^* E_i
\end{equation}
where $\pi$ is the projection from $K_X$ to $X$. Proposition 4.1
of Bridgeland \cite{BridgeTStruct} states that the functor
\begin{equation}
\label{kxequiv}
\Hom(\widetilde{T},\cdot) : \mcD^b(K_X) \longrightarrow
\mcD^b(B - \mathrm{Mod})
\end{equation}
is an equivalence of triangulated categories.

A crucial element of the proof of this theorem is that, for any
line bundle $L$ over a space $X$ with projection $\pi$, we have
\begin{equation}
\label{oproj}
\pi_*(\mcO_L) = \bigoplus_{p\le 0} L^p
\end{equation}
where we also denote by $L$ the sheaf of sections of the line
bundle $L$.

The degree zero part of $B$ contains the idempotents $e_i =
\prod_{k\in\BZ} \mathrm{id}_{E_{i+nk}}$. Thus, we can define the
projective modules $P_i = Be_i$ which are the images of the
objects $\pi^* E_i$. There are also simple modules $T_i$ such
that $\mathrm{dim}(e_i T_j) = \delta_{ij}$. If we let $s$ be the zero
section of $K_X$, then the object $s_*(S_j) = s_*(F_{n-1-j}[j])$ is
mapped to the $T_j$.

Given the equivalence of categories, it is clear by the spectral
sequence of the previous section that the quivers derived there
are the quivers from this algebra. It would be interesting to
use this to prove the conjecture about the superpotential from
the previous section.

\section{Undoing the orbifold}\label{sec:algo}

Now that we see how to derive the quiver for the total space
$K_X$, we would like to interpret it as an orbifold and undo that
orbifold. In this section, we will show how to do so
algorithmically. We will somewhat justify these manipulations in
proceeding sections.

Let us assume we are on a 2-fold, $X$, with Fano index $I$. Let us
denote by $k$ the line bundle such that $k^I = K_X$.
The equivalence (\ref{kxequiv}) of the previous section follows from the
fact (\ref{oproj}) which holds for any line bundle, $L$, and the fact that
the action of tensoring with $K_X$ preserves the helix.  As we
would like to describe the total space of the line bundle
$k$, we would like to have a helix that is preserved by tensoring
with $k$. If this were true, it would follow from the arguments
in \cite{BridgeTStruct} that the invariant portion of the helix
algebra under this action would have a derived category of
modules equivalent to the derived category of coherent sheaves on
the total space of $k$.

We now make the following:

\begin{conj} Given a Fano K\"{a}hler-Einstein 2-fold, $X$, with Fano
index, $I$, and $k$ such that $k^I = K_X$, there always exists a helix,
$E_i$, of length $n$, such that $E_i = E_{i+n/I} \otimes k$.
\end{conj}

In the smooth case, this conjecture is trivial. For $\PP{2}$, we
have $K = \mcO(-3)$, $I = 3$ and $k = \mcO(-1)$. The helix
$(\mcO(i))$ is invariant under tensoring by $\mcO(-1)$. For
$\PP{1}\times\PP{1}$, we have $K = \mcO(-2,-2)$, $I = 2$ and $k =
\mcO(-1,-1)$. The helix (\ref{p1p1hel}) is invariant under
tensoring with $\mcO(-1,-1)$. We should note that in both cases
there exist other helices that do not respect this action. For
example, on $\PP{1}\times\PP{1}$, the helix generated by the
exceptional collection $(\mcO,\mcO(0,1),\mcO(1,0),\mcO(1,1))$
is not invariant. The quiver obtained from this collection is
not, then, an orbifold although it is a perfectly legitimate
description of the gauge theory on $K_{\PP{1}\times\PP{1}}$.
The two quivers should be related by Seiberg duality.

We can provide some evidence for this conjecture in a more
general context by the following. Consider an exceptional
invertible sheaf (line bundle), $A$. Then, clearly, $\Ext^i(A,A) \cong
\Ext^i(A\otimes k^a, A\otimes k^a) = 0$ for all $a$ and $i \neq 0$.
In addition, we have $\Ext^i(A\otimes k^{-a},A) = H^i(k^{a})$.
Now, $k^{a}$ is
a negative line bundle, so this vanishes for $i<2$ by the
Kodaira vanishing theorem. Furthermore, we can apply Serre
duality and, using the fact that $k^{-a} \otimes K_X = k^{I-a}$
is also negative  for $a<I$, we see that this vanishes for $i=2$
also. Finally, $\Ext^i(A,A\otimes k^{-a}) =
H^i(k^{-a})$. Since $k^{-a} = k^{-I-a} \otimes K_x$ and
$k^{-I-a}$ is positive, this Ext group also vanishes for $i >0$.
From this, we can conclude that $(A,A\otimes k^{-1},\dots,A\otimes
k^{-I+1})$ is a strong exceptional collection.

Given that the number of elements in a full exceptional
collection is equal to the rank of the Grothendieck group, this
conjecture implies that the Fano index divides this rank. Again,
this is trivially true in the smooth case. We hope that something
like this holds in the orbifold case.

\begin{figure}
\centerline{
\xymatrix{
\bullet \ar@(r,ur)[] \ar@(ul,l)[] \ar@(dl,dr)[] }
}
\caption{The undone orbifold quiver for $\PP{2}$.}
\label{p2oquiv}
\end{figure}
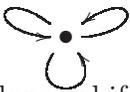

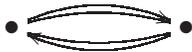
\begin{figure}
\centerline{
\xymatrix{
{\bullet} \ar@/^2mm/[rr] \ar@/^3mm/[rr] & &
{\bullet} \ar@/^2mm/[ll] \ar@/^3mm/[ll] }}
\caption{The undone quiver for $\PP{1} \times \PP{1}$.}
\label{p1p1oquiv}
\end{figure}

Now we can see how to undo the orbifold of the quiver. The action
of $k$ on the helix gives rise to an action on the quiver. In
particular, the nodes of the quiver correspond to the elements in
the exceptional collection. Thus, in the case of $\PP{2}$, the
action is given by rotating figure \ref{p2cquiv} by 120
degrees. For $\PP{1}\times\PP{1}$, figure \ref{p1p1cquiv} is
rotated 180 degrees. By restricting to the invariant parts, we
find the quivers in figures \ref{p2oquiv} and \ref{p1p1oquiv}.
We will make this precise in section \ref{sec:obtainquiver}.

For $\PP{2}$, the superpotential (\ref{p2super}) is invariant
under the action and thus descends to the undone orbifold. As
the cone $\mcO(-1)$ over $\PP{2}$ is just $\MC{3}$ blown up
at the origin, we should obtain $\mathcal{N} = 4$ SYM. To see
that we do, notice that three lines in the quiver correspond
to the three chiral multiplets in $\mathcal{N} = 4$ SYM. The
superpotential gives the usual $[X_i,X_j]^2$ term in the action.

For $\PP{1}\times\PP{1}$, undoing the orbifold identifies
$a$ with $c$ and $b$ with $d$ in (\ref{p1p1super}). This gives us
the following superpotential
\begin{equation}
W = \epsilon^{ik}\epsilon^{lj} a_l b_i a_j b_k\ .
\end{equation}
The cone $\mcO(-1,-1)$ over $\PP{1}\times\PP{1}$ is a resolution of
the cone cut out by $x^2 + y^2 + z^2 + w^2 = 0$ in $\MC{4}$. This cone, 
termed the conifold, was first investigated in the context of AdS/CFT
in \cite{Klebanov:1998hh,Morrison:1998cs}. The
quiver and superpotential obtained here are exactly those derived
there.

\section{Orbifolds of $\BC^*$ bundles}\label{sec:cstar}

In this section, we will show how the above procedure can be
implemented in terms of orbifolds. D-branes on an orbifold,
$Y/G$ for $G$ some finite group, are described by the
equivariant derived category, $\mcD^b_G(Y)$, \cite{Aspinwall:2004jr}.
For simplicity, we will assume $G$ is Abelian. Then, this
category admits an action by the group $G$, the quantum
symmetry, and if we look at the orbits of the action, we
recover the derived category of the original space $\mcD^b(Y)$.
This is the essence of the procedure in the previous section.

There is a barrier to implementing this proposal, however. Given
a line bundle $L$ over $Y$, the total space $L^n$ is \textit{not}
a $\BZ_n$ orbifold of $L$. In this situation,
$\BZ_n$ acts on the fibers by multiplication by $e^{2i\pi/n}$,
but this action preserves the origin. Thus, the quotient space is
singular in codimension one. Later, we will use the McKay
correspondence of \cite{BrKiRe} to overcome this difficulty, but
before tackling that proposition, let us warm up by simply
removing the origin and replacing our line bundles with $\BC^*$
bundles.

$\BC^*$ bundles are particularly relevant for AdS/CFT because
the 5-fold, denoted $M_5$ above, is the total space of a
circle bundle with Euler class, $e$. This bundle is a deformation
retract of the total space of the $\BC^*$ bundle also characterized by
the two-form $e$.

In order to postpone the introduction of
equivariant derived categories, we will deal instead with
equivariant fiber bundles which are ordinary bundles along with a
lift of the orbifold action on the base to the total space of the
bundle. For $x, gx \in Y$ and $g \in G$, this gives an identification of
the points in the fibers $F_x$ and $F_{gx}$. If the fiber is a vector
space, $F$, we can take a representation of $G$, $r : r(g) \in GL(F)$. With
this in hand, we can modify the above identification to $F_x
\rightarrow r(g)F_{gx}$. This gives an action of representations
on equivariant vector bundles. In the case of line bundles, the
one dimensional representations form a group under the tensor
product that, for $G$ abelian, is the same as $G$. Choosing such
an identification gives an action of $G$ on the set of
equivariant line bundles. Note that this is not an automorphism
of the equivariant Picard group.

Let us assume that $G$ acts freely on $Y$. Then, the set of
equivariant line bundles is $H^2_G(Y,\BZ) \cong H^2(Y/G,\BZ)$. We
would like to identify the above action on the second cohomology
of $Y/G$. Let us assume that $Y$ is simply connected. Then, it is
a standard fact that $\pi_1(Y/G) \cong H_1(Y/G,\BZ) \cong G$ as
$G$ is Abelian. The universal coefficient theorem then tells us
that $H^2_\mathrm{tors}(Y/G,\BZ) \cong G$. Thus, we have a $G$
action on $H^2(Y/G,\BZ)$ given by addition of its torsion
elements which we have identified with $G$. These torsion
elements correspond to certain line bundles over $Y$, and the
action is simply the usual tensor product of line bundles.

Now, let us specialize to the case at hand where $Y$ is the
total space of a $\BC^*$ bundle over $X$. This $\BC^*$ bundle is
characterized by its Euler class $e$. There is also a
corresponding line bundle over $X$, which we denote as $E$, which has
first Chern class $e$. Let $\pi$ be the projection from $Y$ to
$X$. Then, the pullback bundle $\pi^*E$ is trivial. This can be
see by constructing a global nonzero section. This implies that
$\pi^*(e) = 0$ in cohomology. In fact, from the the Gysin sequence 
and the fact that $X$ is
simply connected, we have $H^2(Y,\BZ)
\cong H^2(X,\BZ) / \BZ e$.  Now, let $X$ have Fano index $I$
and let $E \cong \pi^*(k^I) \cong \pi^*(K_X).$ In our usual abuse of
notation, we will use the same symbols for line bundles and
their first Chern classes. Then we have
\begin{equation}
H^2(Y,Z) \cong H^2(X,Z)/\pi^*(K_X)\BZ
\end{equation}
and $I\pi^*(k) = 0 \in H^2(Y,Z)$.

For $\PP{2}$, we have $M^5 = S^5/\BZ_3$ and $H^2(S^5/\BZ_3,\BZ) =
\BZ_3$. The second cohomology consists of the torsion elements
$\{0,\pi^*(k),2\pi^*(k)\}$ with $3\pi^*(k)=0$. These correspond precisely to the line
bundles that form the exceptional collection
$(\mcO(-2),\mcO(-1),\mcO)$. The action on the cohomology given by
adding the form $\pi^*(k)$ gives precisely the action on the nodes of
the quiver in the previous section. This lifts to an autoequivalence
of the derived category given by tensoring with $\mcO(-1)$.

For $\PP{1}\times\PP{1}$, we have $M^5 = T^{11}/\BZ_2$, and
$H^2(T^{11}/\BZ_2,\BZ) = \BZ \oplus \BZ_2$. If we denote by $a$ and
$b$ the generators of $H^2(\PP{1}\times\PP{1},\BZ) = \BZ \oplus
\BZ$, we have $2\pi^*(a + b) = 0$ on $T^{11}/\BZ_2$.
Thus, the action on cohomology is given by the addition of
$\pi^*(a + b)$, or tensoring with $\mcO(-1,-1)$. Again, this gives
the previously obtained action on the exceptional collection (\ref{p1p1exc}).

\section{The McKay correspondence}\label{sec:mckay}

We would now like to be able to extend this discussion to the
total space of the \textit{line} bundle $K_X$ over $X$. The goal
is to obtain an autoequivalence $\mathcal{G}$ of $\mcD^b(K_X)$ such
that $\mathcal{G}^I = \mathrm{id}$ where $I$ is, as usual, the
Fano index of $X$. As stated above, the problem is that the
$\BZ_I$ orbifold of the total space of $k$ is \textit{not} $K_X$.
However, as we will see, it is possible to resolve the
(conical) singularity of $k/\BZ_I$ to obtain $K_X$. This will
allow us, through a theorem of Bridgeland, King and Reid
\cite{BrKiRe}, to identify $\mcD^b_{\BZ_I}(k)$ with $\mcD^b(K_X)$.
Unfortunately, we have not been able to use this theorem to
obtain the proper autoequivalence. In the next section, we will
use some further results of Bridgeland \cite{BridgeTStruct} to
conjecturally identify this autoequivalence which corresponds to the quantum
symmetry of the orbifold.

Let us briefly introduce the equivariant derived category. For
further details, see \cite{BrKiRe}. Let $Y$ be a space with a $G$
action. An equivariant sheaf on $Y$ is a sheaf, $\mcF$ on
$Y$, along with a set of maps $f_g: \mcF \rightarrow g^*\mcF$
that satisfy $f_{hg} = g^*(f_h) f_g$ and $f_1 = \textit{id}$.
There is an action of $G$ on $\Hom(\mcF,\mcG)$ for $\mcF$ and
$\mcG$ equivariant sheaves. Restricting to the invariant part, we
obtain $G\dHom(\mcF,\mcG)$. From this, one can obtain the Abelian
category $Coh^G(Y)$ and the usual derived categories including
$\mcD^b_G(Y)$. All the usual functors such as $G$-Ext and pushforwards
and pullbacks for equivariant maps exist and satisfy the usual
relations. As with the equivariant line bundles above, there is also an
action of a representation of $G$ on an equivariant sheaf by
tensor product. 

In order to state the McKay correspondence of \cite{BrKiRe} we
need one final ingredient, the Hilbert scheme of $G$-clusters on
$Y$, $G$-Hilb $Y$. In general, this is a complicated object which we will not
define. Broadly
speaking, however, we can look at the space of subschemes (think
submanifolds) of some space, $Y$. For example, we can look at the
space of sets of $n$ points in $Y$ where the points are
considered indistinguishable. This is, in general, a singular
space when two points come together. The Hilbert scheme is a
crepant resolution of the singular points of this space. The
Hilbert scheme of $G$-clusters, then, is a resolution of the
space of $G$-clusters which are, in essence, orbits
of the $G$ action on $Y$. Thus, $G$-Hilb $Y$ is a resolution of
$Y/G$.

With all this in hand, Bridgeland, King and Reid proved that
\begin{equation}
\label{mckay}
\mcD_G(Y) \cong \mcD(G\mbox{-Hilb\ }Y)\ .
\end{equation}
More properly, we should replace $G$-Hilb $Y$ by its irreducible
component containing the free orbits. There are also some further
technicalities which we will ignore.

We can now specialize to our situation. Let $X$ and $I$ be as before,
and let $Y = k$ with the $G=\BZ_I$ action given by rotation on
the fibers. We would like to determine $G$-Hilb $Y$. The first
thing to note is that, as the action of the group preserves the
fiber, we can reduce the question to determining $G$-Hilb
$\BC$. This can be embedded in Hilb$^{I}$ $\BC$. However, Hilbert
schemes of points in dimension one are known to be isomorphic
to the symmetric product $\MC{I}/S_I$
which is nonsingular in this case. Elements in this space can be written as
sums of complex numbers, \ie $a + 2b + c \in$ Hilb$^4$ $\BC$.
The action of $G$ on $\BC$ lifts to an action on $\MC{I}/S_I$,
and $G$-Hilb is embedded as the $G$ invariants points. It is not hard
to see that these are given by $\sum_{i=0}^I g_i a$ for $a
\in \BC$ and $g_i$ ranging over all the elements in $G$. This
gives a map from $\BC$ to $G$-Hilb $\BC$. Since, off the zero
section, the space $Y/G$ is nonsingular and isomorphic to the
total space $K_X$ off the zero section, we see that $\BZ_I$-Hilb
$k \cong K_X$. Thus, we have the hoped for automorphism
\begin{equation}
\label{zmckay}
\mcD_{\BZ_I}(k) \cong \mcD(K_X)\ .
\end{equation}

This implies that there should be an action of the group $\BZ_I$
on the triangulated category $\mcD(K_X)$, but it appears to be
difficult to determine this directly from the work of Bridgeland,
King and Reid. In the next section, we will conjecture the
correct autoequivalence. As we can embed the $\MC{*}$ bundles of
the previous section into the line bundles of this section, we
expect that, off the zero section, this should reduce to the
autoequivalence discovered there.

\section{The orbifold monodromy}\label{sec:monod}

\subsection{Spherical objects and twist functors}

In order to formulate our conjecture about the orbifold
monodromy, we must introduce the notions of spherical objects and
twist functors due to Seidel and Thomas \cite{SeiThom}. Let $Z$ be
an algebraic variety of dimension $n$. An object in $S \in \mcD(Z)$ is
called \textit{spherical} if
\begin{equation}
\Hom(S,S[k]) = \begin{cases}
              \BC &\text{if $k = 0$ or $n$} \\
	      0   &\text{otherwise.}
              \end{cases}
\end{equation}

An easy source of spherical objects follows from the relation
(\ref{extreln}). This implies that, for any exceptional object,
$E \in \mcD(X)$, $s_*(E) \in \mcD(K_X)$ is spherical where $s$ is the
zero section of $K_X$. Given an exceptional collection, we will
associate a collection of spherical objects as the pushforwards
of the dual collection. Recall that these correspond to the
simple modules of the completed quivers.

Seidel and Thomas then define a \textit{twist functor} $\mcT_S$
which is the autoequivalence of the derived category which
completes the following triangle for all $F \in \mcD(Z)$
\begin{equation}
\Hom(S,F) \otimes S \rightarrow F \rightarrow \mcT_S(F)\ .
\end{equation}
Given a
collection of spherical objects, one can define an action of the
braid group by these autoequivalences of the derived category. This is
further investigated in \cite{BridgeTStruct}.

The twist functor can be written in terms of a Fourier-Mukai
transform\footnote{For an introduction to Fourier-Mukai
transforms, see \cite{Aspinwall:2004jr,Andreas:2004uf}.} as follows.
Recall that for an autoequivalence, the Fourier-Mukai transform
is specified by its kernel, an element in $\mcD(Z \times Z)$.
Let $S \in \mcD(Z)$ be spherical. Then, the kernel that gives
the twist functor is
\begin{equation}
S^{\vee} \boxtimes S \rightarrow \mcO_{\Delta Z}
\end{equation}
where $S^{\vee} = \mathbb{R}Hom(S,\mcO_Z)$ is the dual object to $S$ in the derived category.
Twist functors generally appear in physics as monodromies around
conifold points in the moduli space. For more information, see
\cite{Aspinwall:2004jr} and references therein.

\subsection{The conjectured monodromy}

We will need the following result of Bridgeland, a slight variation of
Proposition 4.9 of \cite{BridgeTStruct}\footnote{For a more careful
exposition, please see the original paper.}. Let $(E_0,\dots,E_{n-1})$
be a simple collection in $\mcD(X)$, let $E_i$, $i \in \BZ$
be the corresponding helix, and let $(S_0,\dots,S_{n-1})$
be the corresponding spherical objects. Then, the spherical
objects for the simple collection $(E_{-1},E_0,\dots,E_{n-2})$
are $\mcT_{S_{n-1}}(S_{n-1},S_0,\dots,S_{n-2})$ where the
twist functor acts separately on each element in the set.

In fact, Bridgeland's proof is only given for the case dim $X = n$.
However, the proof can be seen to hold in general using the
definition of a helix given in section \ref{sec:helmut}.



Now, let us define the autoequivalence,
\begin{equation}
\label{orbauto}
\mcG = \mcT_{S_{n/I-1}} \dots \mcT_{S_0} \Ten_{k^{-1}} 
\end{equation}
where we act from right to left and $\Ten_{k^{-1}}$
denotes the autoequivalence given by tensoring with the
invertible sheaf/line bundle $k^{-1}$. Notice that, off the zero
section, this is precisely the inverse of autoequivalence found for the
$\MC{*}$ bundles of section \ref{sec:cstar}. With a little work, it can be
seen to follow from Bridgeland's theorem that this autoequivalence gives
a cyclic shift in the simple objects. In other words, it takes the collection
$(S_0,\dots,S_{n-1})$ to $(S_{n/I},\dots,S_{n-1},S_0,\dots,S_{n/I-1})$. 

Physically, we know that these spherical objects correspond
to fractional branes, and orbifold monodromies permute the simple
branes. Furthermore, the $I$th power of our autoequivalence
preserves the simples. This motivates us to
\begin{conj}
The autoequivalence of $\mcD(K_X)$ that corresponds to
the action of $\BZ_I$ on $\mcD_{\BZ_I}(k)$ is given by the functor
$\mcG$ in equation (\ref{orbauto}). Furthermore, this corresponds to
the monodromy about the
orbifold point in the moduli space for $K_X$.
\end{conj}
The second point needs some elaboration. In fact, in a
multiparameter moduli space such as that for $\PP{1} \times
\PP{1}$, the orbifold point is of codimension greater than one,
so there is no obvious notion of an orbifold monodromy. One can,
however, take a curve in the moduli space which intersects with
the orbifold point and take the monodromy constrained to that
curve. This may be the proper interpretation of the monodromy
obtained here.

Now, let us apply this to our two examples. For $\PP{2}$, we have
\begin{equation}
\mcG = \mcT_{s_*(\mcO)}\Ten_{\mcO(1)} 
\end{equation}
where, by $\mcO(n)$, we mean the pullback sheaf
$\pi^*(\mcO_\PP{2}(n))$.
This is precisely the well-known orbifold monodromy for
$\MC{3}/\BZ_3$ (see, for example \cite{Aspinwall:2004jr}).
For $\PP{1}\times\PP{1}$, we obtain
\begin{equation}
\label{p1p1autoeq}
\mcG = \mcT_{s_*(\mcO(-1,0))}\mcT_{s_*(\mcO)}\Ten_{\mcO(1,1)}\ .
\end{equation}
We show in the appendix that this squares to the identity.

\subsection{Obtaining the quiver}\label{sec:obtainquiver}

With the action of the orbifold group in hand, we can finally
make rigorous the manipulations of section \ref{sec:algo}. Recall
that the quiver can be determined from the $\Ext$ groups between
its simple representations (\ref{extmean}). We will use our knowledge of the
$G$-Ext groups in the equivariant derived category to obtain those
in the original derived category, thus determining the undone
quiver.

The needed relationship is Lemma 4.1 of \cite{BrKiRe} which we
quote here. Let $E$ and $F$ be $G$-sheaves on $X$. Then, as a
representation of $G$, we have a direct sum decomposition
\begin{equation}
\label{equihoms}
\Hom_X(E,F) = \bigoplus_{i=0}^k G\dHom_X(E\otimes \rho_i, F)
\otimes \rho_i
\end{equation}
over the irreducible representations of $G$, $\{\rho_0,\dots,\rho_k\}$.
It is straightforward to see that this holds for Ext groups as
well.
 
For us, $G$ is the cyclic group $\BZ_I$, and the representation
ring is generated by a single representation. We identify the action of
the generating representation
with the autoequivalence conjectured above which permutes
the simples. This choice is, of course, not unique.
We can now decompose the simples into orbits under
the permutation and, choosing one representative from each orbit,
compute the Ext groups using (\ref{equihoms}).

For example, for $\PP{2}$, we have three simple representations
$S_i$ upon which the group $\BZ_3$ acts as $S_i \rightarrow S_{i +
1\mod3}$. There is only one orbit, and we must only compute one
set of Ext groups:
\begin{equation}
\Ext^i(S,S) = G\dExt^i(S_0,S_0) \oplus G\dExt^i(S_0,S_1)
	\oplus G\dExt^i(S_0,S_2)\ .
\end{equation}
By the McKay correspondence, we can identify the equivariant
Exts with the Exts of the quiver $K_\PP{2}$ in figure
\ref{p2cquiv}. The three $\Ext^1$s from $S_0$ to $S_1$ give the
three arrows in the quiver of figure \ref{p2oquiv}. The three
$\Ext^2$s from $S_0$ to $S_2$ and the $\Ext^3$ from $S_0$ to
$S_0$ give the relations.

Following the same procedure for $\PP{1}\times\PP{1}$ yields the
quiver of figure \ref{p1p1oquiv}.

\section{Acknowledgements}

I am greatly indebted to Tom Bridgeland for providing me a copy of
\cite{BridgeTStruct} and for many e-mail conversations. I would like to thank
Jacques Distler, Gavril Farkas and Chris Herzog for their assistance on
various aspects of this work. I would also like to thank David Ben-Zvi,
Dan Freed and Uday Varadarajan for useful conversations. This material
is based upon work supported by the National Science Foundation under
Grant No. PHY-0071512, and with grant support from the US Navy, Office
of Naval Research, Grant Nos. N00014-03-1-0639 and N00014-04-1-0336,
Quantum Optics Initiative.

\appendix

\section{Squaring the autoequivalence for $\PP{1}\times\PP{1}$}

In this section, we will square the autoequivalence
(\ref{p1p1autoeq}). The calculation uses similar techniques to those
in \cite{Aspinwall:2004jr}, section 7.3.5. Let $X = \PP{1}\times\PP{1}$
and $Y = K_X$ with projection $\pi$ and zero section $s$ as before. Let
$\mcO = s_*(\mcO_X)$ where the latter is the structure sheaf on $X$.
Furthermore, let $\mathcal{A}(n) = \mathcal{A}\otimes \pi^*\mcO_X(n)$
for any sheaf $\mathcal{A}$ on $Y$. Finally, for an object in the
derived category $\mathcal{E}$, we denote its dual $\mathcal{E}^\vee =
\mathbb{R}Hom(\mathcal{E},\mathcal{O}_Y)$.

As a Fourier-Mukai transform, the autoequivalence
(\ref{p1p1autoeq}) has the following kernel:
\begin{equation}
\mcO^\vee(0,1)\boxtimes \mcO(-1,0) \longrightarrow
\mcO^\vee(1,1) \boxtimes \mcO \longrightarrow
\mcO_\Delta(1,1)
\end{equation}
where, in $\mcO_\Delta(1,1)$, we pull back by the diagonal map.

Squaring this, we obtain
\begin{multline}
\mcO(-1,0)^\vee \boxtimes \mcO(-1,0) \longrightarrow
\mcO(0,-1)^\vee \boxtimes \mcO(0,1) \oplus \mcO(-2,0)^\vee
\boxtimes \mcO \\ \longrightarrow
\mcO(-1,-1)^\vee \boxtimes \mcO(1,1) \longrightarrow
\mcO_\Delta(2,2)\ .
\end{multline}
We can write this as $Cone(\mcA \longrightarrow
\mcO_\Delta(2,2))$ where $\mcA$ consists of the first three terms
in the sequence.

Because $X$ is the zero section of the bundle $Y$, we have the
following exact sequence:
\begin{equation}
\label{zeroexact}
0\longrightarrow \mcO_Y(2,2) \longrightarrow \mcO_Y
\longrightarrow \mcO \longrightarrow 0\ .
\end{equation}
From this, we can determine that $\mcO(a,b)^\vee =
\mcO(-2-a,-2-b)[-1]$. Substituting this into $\mcA$, we obtain
\begin{multline}
\mcA[1] = \mcO(-1,-2) \boxtimes \mcO(-1,0) \longrightarrow
\mcO(-2,-1) \boxtimes \mcO(0,1) \oplus \mcO(0,-2) \boxtimes
\mcO \\
\longrightarrow \mcO(-1,-1) \boxtimes \mcO(1,1)
\end{multline}
which is the pushforward of a sequence on $X$. If we conjugate
with the action of tensor products with $\mcO(1,1)$ we obtain
\begin{equation}
\mcO(0,-1) \boxtimes \mcO(2,1)^\vee \longrightarrow
\mcO(-1,0) \boxtimes \mcO(1,0)^\vee \oplus \mcO(1,-1) \boxtimes
\mcO(1,1)^\vee \longrightarrow \mcO \boxtimes \mcO^\vee
\end{equation}
where we have used that $\mcO(a,b)^\vee = \mcO(-a,-b)$ on $X$.
Finally, it follows from theorem 4.4.3 of \cite{GoroMov} that this
sequence is a resolution of the diagonal. The conjugation by
$\mcO(1,1)$ has no effect on the identity, giving
$\mathcal{A} = \mcO_{\Delta X}[-1]$. Then, using the exact
sequence (\ref{zeroexact}) we can write the total action as
\begin{equation}
Cone\left((\mcO_{\Delta Y} (2,2) \longrightarrow \mcO_{\Delta
Y})[-1] \longrightarrow \mcO_{\Delta Y}(2,2)\right) =
\mcO_{\Delta Y}
\end{equation}
This proves that the autoequivalence squares to the identity.

\bibliographystyle{utphys}
\bibliography{thebib}

\end{document}